\documentstyle[12pt]{article}

\newcommand\half{{\scriptstyle{\frac{1}{2}}}}

\newcommand\cE{{\cal E}}

\newcommand\cH{{\cal H}}
\newcommand\cL{{\cal L}}
\newcommand\cV{{\cal V}}
\newcommand\cW{{\cal W}}
\newcommand\cM{{\cal M}}

\newcommand{\bR}{{\bf R}}
\newcommand{\bT}{{\bf T}}
\newcommand{\bZ}{{\bf Z}}

\newcommand\const{\mathop{\rm const}\nolimits}

\def\vc{{\vec c}}
\def\vE{{\vec E}}
\def\vg{{\vec k}}

\def\vp{{\vec p}}
\def\vP{{\vec P}}
\def\vq{{\vec q}}
\def\vQ{{\vec Q}}
\def\vr{{\vec r}}
\def\vR{{\vec R}}
\def\vv{{\vec v}}

\def\v0{{\vec 0}}

\begin{document}

\setlength{\baselineskip}{17pt}

\title{The exotic Galilei group\\
and \\
the ``Peierls substitution''
\\
}

\author{
C.~Duval
\\
Centre de Physique Th\'eorique, CNRS\\
Luminy, Case 907\\
F-13288 MARSEILLE Cedex 9 (France)
\\
and
\\
P.~A.~Horv\'athy
\\
Laboratoire de Math\'ematiques et de Physique Th\'eorique\\
Universit\'e de Tours\\
Parc de Grandmont\\
F-37200 TOURS (France)
}

\date{\today}

\maketitle

\thispagestyle{empty}

\begin{abstract}

Taking advantage of the two-parameter central extension of the	planar
Galilei group, we construct a
non relati\-vis\-tic particle model in the plane. Owing to
the extra structure, the coordinates do not commute.
Our model can be viewed as the
non-relativistic counterpart of the relativistic anyon considered before by
Jackiw and Nair. For a particle moving in a magnetic field
perpendicular to the plane, the two parameters combine with the magnetic
field to provide an effective mass. For vanishing effective mass the phase
space
admits a two-dimensional reduction, which represents the condensation to
collective ``Hall'' motions, and justifies the rule called  ``Peierls
substitution''. Quantization yields the wave functions proposed
by Laughlin to describe the Fractional Quantum Hall Effect.
\end{abstract}


\newpage

\section{Introduction}

The  rule called  ``Peierls substitution'' \cite{Peierls} says that a charged
particle in the plane subject to a strong  magnetic field $B$ and to a weak
electric potential $V(x,y)$ will stay in the lowest Landau level, so that its
energy is approximately $E=eB/(2m)+\epsilon$, where $\epsilon$ is an
eigenvalue of the potential $eV(X,Y)$ alone. The operators $X$ and $Y$
satisfy, however, the anomalous commutation relation
\begin{equation}
\big[X,Y\big]=\frac{i}{eB}.
\label{anomcomm}
\end{equation}

Similar ideas emerged, more recently, in the context of the Fractional Quantum
Hall Effect~\cite{GJ}, where it is argued~\cite{LAUGH} that {\sl the system
condensates into a collective ground state}. This ``new state of matter'' is
furthermore represented by the ``Laughlin'' wave functions~(\ref{BargmannFock})
below, which all belong to the lowest Landau level~\cite{QHE}.

Dunne, Jackiw, and Trugenberger~\cite{DJT} justify the Peierls rule by
considering the $m\to0$ limit, reducing the classical phase space from four
to two dimensions, parametrized by non-commuting coordinates $X$ and $Y$,
whereas the potential $V(X,Y)$ becomes an effective Hamiltonian.
While this yields the essential features of the Peierls substitution,
it has
the disadvantage that the divergent ground state energy $eB/(2m)$ has
to be removed by hand.
In this Letter,
we derive a similar model from {\sl first principles}, without
resorting to such an unphysical limit.

First we construct, following Souriau~\cite{SSD}, a model for a
non-relativistic particle in the plane associated with the
two-parameter central extension \cite{LL,centralex,BGGK,LSZ}
of the Galilei group. Our model, parametrized by the mass, $m$, and a new
invariant,
$\kappa$, turns out to be the non-relativistic limit of the relativistic
anyon model
of Jackiw and Nair~\cite{JN}.

For a free particle the usual equations of motions hold unchanged
and~$\kappa$ only
contributes to the conserved quantities, (\ref{momentmap}). More
importantly, it
yields non-commuting position coordinates, see below. Minimal coupling to an
external gauge field unveils, however, new and interesting phenomena, which
seem to
have escaped attention so far. The interplay between the internal structure
associated with $\kappa$ and the external magnetic field $B$ yields, in
fact, an
effective mass $m^*$. For vanishing effective (rather than real) mass, we
get some
curiously simple motions, which satisfy a kind of {\sl generalized Hall law},
Eq.~(\ref{HallLaw}) below. For a constant electric field the usual cycloidal
motions degenerate to a pure drift of their guiding centers alone. Such motions
form a two-dimensional submanifold of the four-dimensional space of motions.
Reduction to this subspace is the classical manifestation of Laughlin's
condensation into a collective motion. Then the quantization of the reduced
model
allows us to recover the Laughlin description.

\goodbreak
\section{Exotic particle in the plane}

First we construct a classical model of our ``exotic'' particle in the plane.
Let us start with the Faddeev-Jackiw framework \cite{FaJa}.
A  mechanical system is described by the classical action
$\displaystyle\int\!\theta$
defined through the ``Lagrange one-form''
$\theta=a_\alpha{}d\xi^\alpha-Hdt$, where $\xi=(\vr,\vv)$ is a point in phase
space.
The Euler-Lagrange equation
is expressed using~$\omega=\half\omega_{\alpha\beta}d\xi^\alpha\wedge
d\xi^\beta$,
the $t=\const$
restriction of the two-form~$d\theta$, as
\begin{equation}
\omega_{\alpha\beta}\dot{\xi}^\beta=\partial_{\xi^\alpha}H.
\label{EL}
\end{equation}

For a system with a first-order Lagrangian $\cL=\cL(\vec{x},\vec{v},t)$,
for example, one can chose in particular $\theta=\cL\,dt$; when $\omega$ is
regular, we get Hamilton's equations. The construction works, however,
under more
general conditions: on the one hand, not all
one-forms~$\theta$ come from a Lagrangian $\cL$ which would only depend
on position, velocity and time~\cite{Hor}. On the other hand, the
two-form~$\omega$ can suffer singularities, necessitating ``Hamiltonian
reduction'',  which amounts to eliminating  some of variables and writing the
reduced one-form using intrinsic canonical coordinates on the reduced
manifold~\cite{FaJa}.

The Faddeev-Jackiw framework is actually equivalent to that of
Souriau~\cite{SSD},
who proposed to describe the dynamics by a closed two-form, $\sigma$, of
constant
rank on the ``{evolution space}'' ${\cV}$ of positions $\vr$, velocities
$\vv$, and time $t$. Then the classical motions are the integral curves of
the null space of~$\sigma$, viz
\begin{equation}
(\dot\vr,\dot\vv,\dot{t})\in\ker\sigma.
\label{Sform}
\end{equation}
Writing $\sigma$ as $\omega-dH\wedge dt$, the Euler-Lagrange
equations~(\ref{EL})
are recovered. Being closed, $\sigma$ is furthermore locally~$d\theta$, showing
that the two approaches are indeed equivalent.

Working with the two-form $\sigma$ is actually more convenient as working with
the one-form~$\theta$. For example,
a symmetry is a transformation which leaves~$\sigma$ invariant,
while the  Lagrange one-form~$\theta$ changes by a total
derivative.

Souriau \cite{SSD} actually goes one step farther, and
(as advocated also by Crnkovic and Witten~\cite{CW}), argues that
the fundamental space to look at is ${\cal M}$, the
{\sl space of solutions of the equations of motion}.
Souriau calls this abstract substitute of the phase space
the {\sl space of motions}.
In our case, ${\cal M}$ is the simply the set of motion curves
in the evolution space~$\cV$.

Our classical  particle model is then constructed as follows. Let us recall
that
the elementary particles correspond to irreducible, unitary representations of
their symmetry groups. According to geometric quantization, though, these
representations are associated with some coadjoint orbits of the symmetry
group~\cite{SSD,KoKir}; the idea of Souriau~\cite{SSD} was to view these
orbits, endowed with their canonical two-forms, as {\sl spaces of motions}.

\goodbreak


Now, as discovered by L\'evy-Leblond~\cite{LL}, the planar Galilei group
admits a two-parame\-ter central extension,
parametrized by two real constants~$m$ and~$\kappa$ (see, e.g.,~\cite{GRIG}).
The new invariant $\kappa$ has the dimension of $\hbar/c^2$. The
coadjoint orbits of the doubly-extended Galilei group coincide with those of
the singly-extended one, but carry
a modified symplectic structure. The interesting ones are those
associated with the mass $m>0$ and $\kappa\neq0$; they are
$\cM=\bR^4$
with coordinates $(v_i)$ and $(q^i)$, endowed with
the noncanonical {\em twisted-in-the-wrong-way} symplectic two-form
\begin{equation}
\omega
=
m\,\,dv_i\wedge dq^i
+
\half \kappa\,\varepsilon_{ij}\,dv^i\wedge{}dv^j.
\label{symplectic}
\end{equation}
Owing to the new term in~(\ref{symplectic}), the Poisson bracket of the
configuration coordinates is nonvanishing, $\{x,y\}=\kappa/m^2$.
For these orbits, the evolution space is $\cV=\cM\times\bR\simeq\bR^5$,
endowed with the two-form
\begin{equation}
\sigma
=
m\,\,dv_i\wedge(dr^i-v^i dt)
+
\half\kappa\,\varepsilon_{ij}\,dv^i\wedge{}dv^j.
\label{presymplectic}
\end{equation}

This two-forms is exact, namely $\sigma=d\theta$ with
$\theta=mv_{i}dr^{i}-\half
m\vert\vv\vert^2dt+\half\kappa\epsilon_{ij}v^{i}dv^{j}$.
However, because of the ``exotic'' contribution,
it is {\sl not} of the form $\cL\,dt$ with a first-order Lagrangian
$\cL$ \cite{Hor};
thus, this model has {\sl no ordinary Lagrangian}.
Both generalized formalisms work nevertheless perfectly,
and we choose to pursue along these lines.
(Let us mention that a Lagrangian could be
constructed---but it would be acceleration-dependent~\cite{LSZ}.)

Most interestingly, the ``exotic'' term
$\half\kappa\,\varepsilon_{ij}\,dv^i\wedge{}dv^j$ in (\ref{symplectic}) has
already
been used, namely to describe relativistic anyons~\cite{JN}; our
presymplectic form
(\ref{presymplectic}) appears to be the non relativistic limit of that in
Ref.~\cite{JN} when their spin, $s$, is identified with our
parameter~$\kappa$. (We believe in fact that our particles
are indeed non-relativistic anyons.)
Group contraction of the (trivially) centrally-extended
Poincar\'e group yields furthermore
the doubly-extended planar Galilei group~\cite{BGGK}.

It is readily seen that the modified two-form~(\ref{presymplectic}) yields
the usual
equations of free motions,
despite the presence of the new invariant~$\kappa$.
The two-form~(\ref{presymplectic}) on $\cV$ flows down to~$\cM$ as $\omega$ in
(\ref{symplectic}) along the projection $(\vr,\vv,t)\to(\vq,\vv)$, where
$\vq=\vr-\vv t$. The space of free motions is hence $\cM$, endowed with the
symplectic form~$\omega$.

For completeness, let us mention that $\sigma$ is invariant with respect to the
natural action of the Galilei group on $\cV$ whose ``moment map''~\cite{SSD}
consists of the conserved quantities
\begin{equation}
\left\{
\begin{array}{rl}
\jmath&= m\,\vr\times\vv+\half \kappa\vert\vv\vert^2,\\
\\
k_{i}&= m(r_i-v_it)+\kappa\,\varepsilon_{ij}v^j,\\
\\
p_i &= mv_i,\\
\\
h&=\half m\vert\vv\vert^2.
\end{array}
\right.
\label{momentmap}
\end{equation}
These same quantities were found before (see \cite{centralex,BGGK}), using
rather different methods.
Let us observe that, owing to the exotic structure,
the angular momentum $\jmath$ and the boosts $\vg$ in (\ref{momentmap})
contain new terms (which are, however, also separately conserved).
By construction, they satisfy the commutation relations
of the doubly-extended planar Galilei group which only differ from the usual
ones in that the boosts no longer commute,
$\{k_i, k_j\}=\kappa\varepsilon_{ij}$, cf.~\cite{LL,centralex,BGGK}.


Let us now put our charged particle into an external electromagnetic field
$F=(\vE, B)$. Applying as in~\cite{SSD} the minimal coupling prescription
$\sigma\to\sigma+eF$, the system is now described by the two-form
\begin{equation}
\sigma
=
(m\,dv_i-{e}E_i dt)\wedge(dr^i-v^idt)
+
\half\kappa\,\varepsilon_{ij}\,dv^i\wedge{}dv^j
+
\half eB\,\varepsilon_{ij}\,dr^i\wedge{}dr^j
\label{exoticTwoForm}
\end{equation}
on the evolution space $\cV$.
It is interesting to note that our two-form
(\ref{exoticTwoForm})---which is again exact if $F$ is exact,
but is in no way Lagrangian---is the non-relativistic limit
of the relativistic expression in Ref.~\cite{CNP}.
A short computation shows that a tangent vector
$(\delta\vr,\delta\vv,\delta{t})$ satisfies
the Euler-Lagrange equations (\ref{Sform}) when
\begin{equation}
\begin{array}{ccc}
\left\{
\begin{array}{l}
\displaystyle
m^*
\delta{}r^i
=
m\left(v^i-\frac{e\kappa}{m^2}\,\varepsilon^i_jE^j\right)\delta{t},\\
\\
\displaystyle
m\,\delta{}v^i=e\left(E^i\delta{t}+B\,\varepsilon^i_j\delta r^j\right),\\
\\
m\,v_i\delta{}v^i=e\,E_i\delta{}r^i,\\
\end{array}
\right.
\quad\hfill
&\hbox{where}\quad\hfill
&m^*=m-\displaystyle\frac{\kappa{}eB}{m}.\hfill
\end{array}
\label{kernel}
\end{equation}

If the effective mass $m^*$ is nonzero, the third equation is
automatically satisfied; the middle one becomes
\begin{equation}
m^*\delta v_i
=e\left(E^i+B\,\varepsilon^i_j{}v^j\right)\delta{t}.
\label{eqmot}
\end{equation}
Thus, for $\kappa\neq0$, the velocity $\delta\vr/\delta{t}$ and the
``momentum'' $\vv$ are {\sl different} (not even parallel); it is the
latter which satisfies the Lorentz equations of motion~(\ref{eqmot}) with
effective mass~$m^*$.

If, however, {\sl the effective mass $m^*$ vanishes}, i.e., when the
magnetic field
$B$ takes the critical (constant) value
\begin{equation}
B=
\frac{m^2}{e\kappa},
\label{theCondition}
\end{equation}
then $\sigma$ suffers singularities.
The curious ``motions'' with instantaneous propagation can be avoided and
we can still have consistent equations of motion, {\sl provided}
$v^i=(e\kappa/m^2)\varepsilon^i_jE^j$. But this latter condition, together
with Eq.~(\ref{theCondition}), astonishingly reads
\begin{equation}
v^i=\frac{1}{B}\,\varepsilon^i_jE^j.
\label{HallLaw}
\end{equation}
This generalized {\sl Hall law} requires that particles move with
``momentum'' $\vv$ perpendi\-cular to the electric field and determined by the
ratio of the (possibly position and time dependent) electric and the
(constant) magnetic fields.

Assume, from now on that, the electric field $\vec{E}=-\vec{\nabla}V$ be
time-independent.
On the three-dimensional submanifold $\cW$ of $\cV$ defined by
Eq.~(\ref{HallLaw}),  the two-form~(\ref{exoticTwoForm}) induces a
well-behaved closed two-form $\sigma_\cW$ of rank~$2$. Upon defining
the new ``position'' variables
\begin{equation}
Q^i = r^i - \frac{mE^i}{eB^2},
\label{Q}
\end{equation}
one readily finds that
\begin{equation}
\sigma_\cW
=
\half eB\,\varepsilon_{ij}\,dQ^i\wedge dQ^j
- dH\wedge{}dt
\label{sigmaW}
\end{equation}
with the (reduced) Hamiltonian
$
H=eV(\vr)+ {m\vert\vec{E}\vert^2}/({2B^2})
$.
The second term, here, represents the drift energy.
The equations of motion  are simply
\begin{equation}
\begin{array}{ccc}
\left\{
\begin{array}{l}
\displaystyle
\dot{Q}^i
=
\frac{1}{B}\,\varepsilon^i_jE^j,\\[10pt]
\dot{H}=0,
\end{array}
\right.
\end{array}
\label{kersigmaW}
\end{equation}
confirming that the Hamiltonian  descends to the reduced space of
motions. The latter is
two-dimensional and endowed with a symplectic two-form, we call~$\Omega$,
inherited from~$\sigma_\cW$.
Easy calculation shows that
$\partial{H}/\partial{Q^i}=-eE_i$, hence
\begin{equation}
H=eV(X,Y)
\label{H}
\end{equation}
where $(X,Y)$ are coordinates on the reduced space of motions, $\cH$,
obtained by integrating the equations of motion (cf. Eq.~(\ref{R}) below).
Note that the drift
energy has been absorbed into $H$ by the redefinition of the position,
Eq. (\ref{Q}).
At last, one finds that the coordinates~$X$ and $Y$ on $\cH$ have anomalous
Poisson
bracket
\begin{equation}
\{X,Y\}=\frac{1}{eB}.
\label{xyanompb}
\end{equation}
In conclusion, we have established via Eqs (\ref{H}) and (\ref{xyanompb})
the classical counterpart of the Peierls rule.
Let us insist that our construction does not rely on any unphysical
limit of the type $m\to0$, rather it uses the new freedom of having a
vanishing effective mass.

\section{Hall motions}

Let us assume henceforth that the electric field $\vec{E}$ is  constant.
The equations of motion are readily solved.
For nonzero effective mass $m^*$, i.e., when the magnetic field does not
take the critical value (\ref{theCondition}), one recovers the usual motion,
composed of uniform rotation (but with modified frequency $eB/m^*$) plus the
drift of the guiding center.

When the magnetic field takes the critical value (\ref{theCondition})
and when the constraint (\ref{HallLaw}) is also satisfied,
velocity and ``momentum'' become the same, $\vv={\delta\vr}/{\delta{t}}$, so
that the constraint (\ref{HallLaw}) requires that all particles move
collectively, according to \dots {\sl Hall's law}~! This is understood by
noting that for  vanishing effective mass
$m^*=0$,
the circular motion degenerates to a point, and we are left with
the uniform drift of the guiding center alone.

The reduced space of motions $\cH$ (we suggestively called the {\sl space of
Hall motions}) can now be described explicitly.
It is parametrized (see Eqs (\ref{Q}) and (\ref{kersigmaW})) by the coordinates
$(X,Y)\equiv(R^i)$ where
\begin{equation}
R^{i}=Q^i-\frac{1}{B}\,\varepsilon^i_jE^j\,t.
\label{R}
\end{equation}

The constraint (\ref{HallLaw}) implies now that $\delta\vec{v}=0$; the induced
presymplectic two-form on the three-dimensional submanifold $\cW$ is hence
simply $eF$. The symplectic structure of the space of Hall motions is therefore
\begin{equation}
\Omega= \half eB\varepsilon_{ij}\,dR^i\wedge dR^j
= eB\,dX\wedge dY.
\label{symplecticHall}
\end{equation}
The coordinates $X$ and $Y$ have therefore the Poisson bracket
(\ref{xyanompb}).

\goodbreak

The symmetries and conserved quantities can now be found.
Firstly, the ordinary space translations $(\vr,\vv,t)\to(\vr+\vc,\vv,t)$
are symmetries for the reduced dynamics, since they act on $\cH$ according to
$\vR\to\vR+\vc$. The associated conserved quantities identified as the
``reduced momenta'' are linear in the position and time; they read
\begin{equation}
\begin{array}{cc}
P_{i}=-eB\varepsilon_{ij}R^j
=-eB\varepsilon_{ij}Q^j-eE_{i}\,t.
\label{redmom}
\end{array}
\end{equation}
(Their conservation can also be checked directly using the Hall law
(\ref{HallLaw})).
The reduced momenta do not commute but have rather the Poisson
bracket of ``magnetic translations'',
\begin{equation}
\big\{P_{X}, P_{Y}\big\}=eB.
\end{equation}

The time translations $(\vr,\vv,t)\to(\vr,\vv,t+\tau)$ act on
$\cH$ according to $R^i\to{}R^i-\varepsilon^i_jE^j\tau/B$, which is
a combination of space translations.
The reduced Hamiltonian is (see (\ref{H}))
\begin{equation}
H=-e\vec{E}\cdot\vec{R}=-e\vec{E}\cdot\vec{r}
\label{redham}
\end{equation}
and is related to the reduced momenta by
$H=-\vec{E}\times\vec{P}/B$.
The remaining Galilean generators $\jmath$ and $\vg$ are plainly broken
by the external fields. (The system admits instead  ``hidden'' symmetries
that will be discussed elsewhere.)

It is amusing to compare the reduced expressions with the
conserved quantities $\vp$ and $h$ associated with these same symmetries
acting on the original (but ``exotic'') evolution space $\cV$ ``before''
reduction. We find
$
\vp=m\vv+\vP
$
and
$
h=\half m\vert\vv\vert^2+eV\equiv\half m\vert\vv\vert^2+H
$,
where the velocity is of course fixed by the Hall law.
Our reduced expressions are hence formally obtained
by the ``$m\to0$ limit'', as advocated in Ref.~\cite{DJT}.

\goodbreak

Our construction here appears as a nice illustration of
 Hamiltonian reduction~\cite{FaJa}.
The restriction to the $t=\const$ phase space
of our two-form~$\sigma$ in~(\ref{exoticTwoForm}) is a closed
two-form, $\omega$. The generic case, $m^*\neq0$, above arises when $\omega$ is
regular, so that the matrix $\omega$ is invertible.
On the other hand, vanishing effective mass, $m^*=0$,
as in (\ref{theCondition}),
means precisely that~$\omega$ is singular. In Faddeev-Jackiw language, our
reduction
amounts to eliminating the velocities by the constraint~(\ref{HallLaw}) to
yield $X$ and $Y$ as conjugate canonical variables
and $H$ as the Hamiltonian, on reduced space. This is seen by
writing~$\sigma_\cW$,
in~(\ref{sigmaW}), as~$d\theta_\cW$ with Lagrange  form
$\theta_\cW=\half eB\,\varepsilon_{ij}\,Q^idQ^j-H(\vv,\vQ)dt$;
note that the~$dv^{i}$ are absent and
the $\vv$ only appear in the Hamiltonian and are determined
by~(\ref{HallLaw}).

\section{Quantization of the Hall motions}

The quantization is
simplified by observing that the space of Hall motions is actually the same
of that
of a one-dimensional harmonic oscillator with cyclotron frequency~$eB/m$. The
standard procedures~\cite{SSD, KoKir} can therefore be applied.

Let us assume that we work on the entire plane and introduce the complex
co\-ordinate
$Z=\sqrt{eB}\big(X+iY)$; the symplectic form~(\ref{symplecticHall}) is then
$\Omega=d\bar{Z}\wedge dZ/(2i)$, hence $\{\bar{Z},Z\}=2i$.
Now $\Omega$ is exact,
$\Omega=d\Theta$ with the choice
$\Theta=\big(\bar{Z}dZ-Zd\bar{Z})/(4i)$
corresponding to the ``symmetric gauge''.
The prequantum line-bundle is therefore trivial; it
carries a connection with covariant derivative
$D=\partial-\frac{1}{4}\bar{Z}$ along $\partial$.
Choosing the antiholomorphic polarization, spanned by $\bar{\partial}$, yields
the  wave ``functions'' as half-forms $\psi(Z,\bar{Z})\sqrt{dZ}$ that are
covariantly constant along the polarization, i.e., such that
$\bar{D}\psi=0$. This
yields
\begin{equation}
\psi(Z,\bar{Z})=f(Z)e^{-\vert{}Z\vert^2/4}
\label{BargmannFock}
\end{equation}
with $f(Z)$ holomorphic, $\bar{\partial}f=0$.
The the inner product is
$\langle{}f, g\rangle=
\int_{\cH}\overline{f(Z)}g(Z)e^{-\vert{}Z\vert^2/2}\,\Omega$.
We recover hence the ``Bargmann-Fock''~\cite{BARG} wave functions proposed by
Laughlin~\cite{LAUGH}, and by Girvin and Jach~\cite{GJ} to explain the FQHE.
These wave functions span a subspace of the Hilbert space of the ``unreduced''
system and, indeed, represent the ground states in the FQHE~\cite{QHE}. (The
details of the projection to the lowest Landau level are not yet completely
clear, though~\cite{Ouvry}.)

The quantum operator associated to the polarization-preserving
classical observables are readily found~\cite{ KoKir}.
For example, the quantum operators~$\hat{Z}$ and $\bar{Z}$ are given as
$\hat{Z}\psi=(-2\bar{D}+Z)\psi$ and
$\widehat{\bar{Z}}\psi=(2D+\bar{Z})\psi$. Acting on the holomorphic
part alone, this
yields for the complex momenta $\widehat{P}=\widehat{Z}$
and $\widehat{\bar{P}}=\widehat{\bar{Z}}$
\begin{equation}
\left\{
\begin{array}{l}
[\widehat{Z}f](Z)=Zf(Z),\\[10pt]
[\widehat{\bar{Z}}f](Z)=2\,\partial{}f(Z).
\end{array}
\right.
\label{momops}
\end{equation}
(See also \cite{GJ}.) Quantization of polarization-preserving observables takes
Poisson
brackets into commutators; in particular, we have
$\half[\widehat{\bar{Z}},\widehat{Z}\,]=1$,
so that $\widehat{Z},\,\widehat{\bar{Z}}$ and the identity
span the Heisenberg algebra, just
like their classical counterparts.

Being a combination of translations, the reduced Hamiltonian
(\ref{redham})---different from the usual quadratic oscillator
Hamiltonian---becomes
$
H=(\bar{\cE}Z+\cE\bar{Z})/(2B)
$,
once we have put $\cE=\sqrt{eB}(E_1+iE_2)$. Its quantum counterpart is found as
$
\hat{H}=
(\bar{\cE}\hat{Z}+\cE\hat{\bar{Z}})/(2B)
$
with~(\ref{momops}).
For an electric field in the $x$ direction, for example,
\begin{equation}
[\widehat{H}f](Z)=a(2\partial+Z)f(Z),
\label{quantredham}
\end{equation}
where $a=\half{}E\sqrt{e/B}$.
(The subtle problem of ordering does not arise here.)
The eigenfunctions of $\hat{H}$ in (\ref{quantredham}) are readily
found as $f(Z)=Ae^{-(Z-Z_{0})^2/4}$ associated with the (real)
eigenvalue  $\epsilon=aZ_0$, cf.~\cite{QHE}.

Thus the Peierls rule is confirmed
also at the quantum  level, for a linear potential.

\goodbreak

\section{Discussion}

In the spirit of Dirac, we believe that ``{\sl it would be surprising if Nature
would not seize the opportunity to use the new invariant $\kappa$.}'' While it
plays  little r\^ole as long as the particle is free,  this invariant
becomes important when the particle is coupled to an external field: albeit
the classical motions are similar to those in the case~$\kappa=0$, it
yields effective terms responsible for the reduction we found here.
This curious interplay between the ``exotic'' structure and the external
magnetic field is  linked to the two-dimensionality of space and to the
Galilean invariance of the theory.
Mathematically, the second extension parameter arises owing to the
commutativity of planar rotations---just like for exotic statistics  of
anyons~\cite{Leinaas}.
The physical origin of $\kappa$ is, perhaps, the band structure. In a solid the
effective mass can be as much as 30 times smaller than that of a free electron.
Our formula~(\ref{theCondition}) could indeed serve to measure the
new invariant~$\kappa$ using the data in the~FQHE.

Had we worked over the two-torus~$\bT^2$ rather than over the whole plane,
pre\-quantiza\-tion would require the integrality condition
$\displaystyle\int\!\Omega=2\pi\hbar N$ for some integer $N$ \cite{SSD,
KoKir}.
The actual meaning of this condition is that the ``Feynman'' factor
\begin{equation}
\exp\left(\frac{i}{\hbar}\int\!\theta\right)
\label{Fefa}
\end{equation}
be well-defined, independently of the choice of the
one-form~$\theta$~\cite{Feynman}.

Representing $\bT^2$ by a rectangle of sides $L_{x}$ and $L_{y}$ would then
imply
the well-known magnetic flux quantization condition
$eBL_{x}L_{y}=2\pi\hbar N$\cite{QHE}, analogous to the
Dirac quantization of monopoles. Furthermore, the
non-simply-connectedness
of the torus implies that the factor (\ref{Fefa}) can
have different inequivalent meanings, labeled by the characters of
$\bZ\times\bZ$, the homotopy group of the two-torus  \cite{SSD, Feynman,
torusQ}.

\section{Acknowledgement}
We are indebted to Prof. R.~Jackiw for acquainting us with the Peierls
substitution and for advice,
and to Profs. J.~Balog, P.~Forg\'acs and Z. Horv\'ath for discussions.
Correspondence with Profs. V.~P.~Nair and ~G.~Dunne
is also acknowledged.


\end{document}